\documentclass{aastex}
\input{epsf.sty}
\newcommand{\etal}{et al.~}
\newcommand{\lsun}{\hbox{L$_{\odot}$}}
\newcommand{\msun}{\hbox{M$_{\odot}$}}

\begin{document}
 \title{Dynamical Masses of T Tauri Stars and Calibration of PMS Evolution}
 \author{M. Simon}
 \affil{Dept. of Physics and Astronomy}
 \affil{State Univ. of New York, Stony Brook, NY 11794-3800, U.S.A.}
 \author{A. Dutrey and S. Guilloteau}
 \affil{Institut de Radio Astronomie Millim\'etrique}
 \affil{300 Rue de la Piscine, F-38406 Saint Martin d'H\`eres, France}

\begin{abstract}

We have used the high sensitivity and resolution of the IRAM
interferometer to produce sub-arcsecond $^{12}$CO J=2-1 images of
9 protoplanetary disks surrounding T\,Tauri stars in the
Taurus-Auriga cloud (7 singles and 2 binaries). The images
demonstrate the disks are in Keplerian rotation around their
central stars. Using the least square fit method described in
Guilloteau \& Dutrey (1998), we derive the disk properties, in
particular its inclination angle and rotation velocity, hence the
dynamical mass. Since the disk mass is usually small, this is a
{\it direct} measurement of the stellar mass. Typically, we reach an 
internal precision of 10\% in the determinations of stellar mass.
The over-all accuracy is limited by the uncertainty in the distance
to a specific star. In a distance independent way, we
compare the derived masses with theoretical tracks of
pre-main-sequence evolution. Combined with the mean distance to
the Taurus region (140~pc), for stars with mass close to 1 \msun, our 
results tend to favor the tracks with cooler photospheres (higher 
masses for a given spectral type). We find that in UZ Tau E the 
disk and the spectroscopic binary orbit appear to have different inclinations.
\end{abstract}

\keywords{Stars: T\,Tauri -- circumstellar matter -- pre-main
sequence -- individual: BP Tau, CY Tau, DL Tau, DM Tau, GG Tau, GM
Aur, LkCa15, MWC 480, UZ Tau
 -- binaries: close -- Radio-lines: stars -- }

\section{Introduction}
Nearly all of our knowledge about the masses and ages of low mass young stars 
comes from their location in the HR diagram relative to theoretical 
calculations of stellar evolution to the main sequence. Despite considerable 
advances over the past 5-10 years in understanding the structure and 
atmospheres of stars of mass M$< 1~$ \msun, comparison of the currently 
available predicted evolutionary paths of young stars
shows obvious differences.  Empirical tests of the  calculations have not been
possible until recently because astronomers have not had independent
measurements of either the mass or age of a young star.  This situation is 
changing rapidly.  One way to test the calculated tracks is to investigate 
whether they yield the same ages for stars expected to be coeval on physical 
grounds.  Hartigan et al. (1994), Casey et al. (1998), and White \etal
(1999) have applied this test to wide binaries, the TY CrA system, and
the GG Tau system, respectively, using tracks available at the time.
Over a mass range extending down to the brown dwarfs, they found significant
differences in the extent to which the several theoretical calculations 
satisfy the coevality requirement.

The capability of mm-wave interferometers to resolve the spectral line 
emission of the outer disks of PMS stars  offers the possibility to map their 
rotation. Since the disk mass is usually very small compared to the 
mass of the star, this provides  the means to measure the stellar mass 
dynamically (e.g. Dutrey \etal 1994, DGS94; Guilloteau and Dutrey 1998, 
GD98).  We use this technique to obtain new measurements of the
mass of 5 single PMS stars and one PMS binary.  Combined with our previous 
measurements of the PMS single GM Aur and the binary 
GG Tau (Dutrey \etal 1998, Guilloteau, Dutrey and Simon 1999) the measured 
stellar masses span the
range $\sim 2$ to 0.5 \msun. We use the results to test the theoretical 
calculations of PMS evolution.  Sect. 2 describes the sample of 
stars and the interferometric observations.  Sect. 3 summarizes the analysis 
and presents the measured masses and related parameters. Sect. 4 compares 
the measured masses with those implied by location of the 
stars in the HR diagram relative to the tracks calculated by D'Antona and 
Mazzitelli (1997, DM97), Baraffe \etal (1998, BCAH), Palla and Stahler 
(1999, PS99), and Siess \etal (2000, SDF).

\section{Sample of Stars and Interferometer Observations}
\subsection{Sample of Stars}
All the stars studied are in Taurus-Auriga (Table 1).  We obtained 
new interferometric observations in the $^{12}$CO J= 2-1 line of the 
singles CY Tau, DL Tau, DM Tau, Lk\,Ca\,15, and MWC\,480 (HD 31648), and 
the spectroscopic binary UZ Tau E (Mathieu \etal 1996). They were selected 
 because earlier interferometric observations (Table 1, Col. 5) indicated 
that they were, or were likely to be, associated with resolvable circumstellar
 disks.   BP Tau was observed because its \textit{HIPPARCOS} distance, 
56$\pm14$~pc (Favata \etal 1998), is very different from the 140 pc average 
distance to the Taurus star forming region (Kenyon \etal 1994), suggesting 
that it is much older than previously thought (see also Bertout \etal 1999). 
Mass measurements of the single GM Aur and the binary GG Tau Aa by $^{12}$CO 
J= 2-1 line interferometric mapping have already been reported  (Dutrey \etal 
1998 and Guilloteau \etal 1999). We included DM Tau in our program  
because new data at 3 times better angular resolution would 
provide an independent check on the method and earlier results (GD98).

Table 1 provides spectral types, luminosities, and effective temperatures
for our sample.  Most of the spectral types and stellar luminosities, L$_*$,  
in Cols. 2 and 3 of Table 1 are taken from Kenyon and Hartmann (1995).  The 
luminosities are the L$_J$ in their Table A4.  All the luminosities in Col. 
3 are evaluated at 140 pc distance.  The actual distances to the individual 
stars are important for our analysis but are presently unknown; we will 
discuss this problem in Sect. 4. The effective temperatures in Col. 4, 
T$_{eff}$, corresponding to the spectral types, for all the stars, are 
obtained using Table A1 of Kenyon and Hartmann (1995) for main sequence
stars; we discuss the spectral-type to T$_{eff}$ conversion further in Sect. 4.

\subsection{Interferometer Observations}
Our IRAM interferometer observations used 5 antennas operating in snapshot mode
(Guilloteau \etal 1992, Dutrey \etal 1996) and were carried out in winter 
1997/1998.  For BP Tau, the full synthesis mapping was obtained in winter 
1998/1999.  We observed all the sources simultaneously at 89.2 
(HCO$^+$ J= 1-0) GHz and 230.5 GHz ($^{12}$CO J= 2-1).  The spectra were 
analyzed using a correlator with one band of
10 MHz centered on the HCO$^+$ J= 1-0 line, one band of 20 MHz centered on the
$^{12}$CO J= 2-1 line, 2 bands of 160 MHz for the 1.3 mm and 3.4 mm continuum,
respectively. The spectral resolution was 0.23 and 0.18 km/s in the narrow 
bands at 3.4 and 1.3mm, respectively. The phase calibrators were  0415+379 
(3C 111) and PKS 0528+134.  The rms phase noise was 8$^\circ$ to 25$^\circ$ 
and 15$^\circ$ to 50$^\circ$ at 3.4 mm and 1.3 mm, respectively, which 
introduced position errors of $<0.1''$. The seeing, estimated from 
observations of the calibrators, was $\sim0.3''$. Baselines up to 400
m provided $\sim 0.7''$ resolution for the 1.3 mm continuum data and 
1.8$''$ at 3.4 mm. The flux scale was checked by observation of MWC349 which 
has been used as calibrator at IRAM since 1996. Its flux is given by 
$S_{\nu} = 0.93\times(\nu/87)^{0.6}$~Jy, following Altenhoff \etal (1981), 
and subsequent measurements of the planets made at IRAM (IRAM Flux report 13).

The GILDAS software package was used to reduce the data. At 1.3 mm, the 
continuum images are the results of summing the lower and the upper side-band 
data. At 3.4 mm, the tuning was purely in the lower side band. The continuum 
and line maps were produced using natural weighting of the visibilities. For 
clarity, tapering to about $1''$ resolution was used on line maps presented 
in Figs.1 and 2, but the full angular resolution was used in the fitting of 
models by the $\chi^2$ technique. We did not subtract the continuum emission 
from the CO map because its contribution is small, but included it in the 
analysis.

\section{Analysis and Results}
The essential inputs are the angularly resolved maps of the CO line 
emission at each
velocity channel and the ``standard'' model of a rotating disk in hydrostatic
equilibrium (e.g. DGS94).  We use the procedure developed by GD98 to derive the
parameters of the model by the method of $\chi^2$ minimization. In order to 
avoid nonlinear effects arising from deconvolution, the $\chi^2$ minimization 
is performed in the {\it u,v} plane.  The derived parameters divide naturally
into those relevant for the determination of the central mass and those 
that pertain mostly to the properties of the disk.  Here, with the exception 
of DM Tau, for which
we summarize the complete analysis, we will concentrate on the parameters 
required to determine the central mass - the disk inclination $i$ and 
position angle $PA$,
$V_\circ$ sin$i$ at $r_\circ$ = 100 AU, the exponent $v$ of the rotation 
curve $V_\circ (r/r_\circ)^{-v}$ and the outer disk radius $R_{out}$.  A paper 
describing the disk parameters (e.g. the temperature radial distribution, 
the $^{12}$CO abundance and spectral line turbulence width) is in preparation.

GD98 demonstrated the $\chi^2$ parameter fitting procedure by applying it to
$^{12}$CO J= 1-0 data for the single star DM Tau. They used D=150 pc for the
distance to DM Tau.  Since we now adopt 140 pc as the reference distance, 
we scale GD98's results to this value. On this basis, GD98's value for 
DM Tau's mass is $(0.47\pm0.06) \times (D/140$pc)~$\msun$ (Table 3).
Fig.~1 shows 7 of the velocity channel maps measured in the $^{12}$CO J=2-1  
line for all the sources in our sample. Fig.~1 also presents velocity 
gradient maps which show on the red and blue shifted sides of the systemic 
velocity, the characteristic CO emission pattern attributable to rotation of a 
tilted disk.

Fig. 1a shows that DM Tau, GM Aur, Lk\,Ca\,15 and MWC\,480 appear to have 
no CO emission from the parent cloud.  UZ Tau E and DL Tau (Fig. 1b) suffer 
some confusion with the CO emission of the cloud.  Confusion is maximum at 
the cloud systemic velocity, which is usually (but not always, see DL Tau) 
the stellar velocity. Hence it preferentially affects the disk emission 
which is elongated along the projected disk minor axis. This results in an 
incorrect estimate of the disk inclination.  In the case of CY Tau, we cannot 
be sure about the level of confusion because of the weakness of its CO 
emission.  For UZ Tau E and DL Tau, the velocity channels hidden by the 
foreground CO cloud have been removed from the $\chi^2$ analysis.  In the 
case of UZ Tau E, 4 out of 36 significant channels were removed. In the case 
of DL Tau, 5 out of 18 significant channels had to be removed because of
confusion. This increased the uncertainty in the disk inclination quite
significantly.

The first step of the model fitting procedure is to measure the 
coordinates of the mm-wave continuum source in order to recenter the 
$u,v$ plane visibilities on it.
Table 2 lists these source coordinates for DM Tau and for the other stars newly
observed for this program. The UZ Tau E spectroscopic binary cannot be 
resolved by our observations; its continuum source coordinates must be 
effectively centered on the binary.

Table 3 compares the results of the full analysis of DM Tau obtained by 
GD98 from their analysis of the $^{12}$CO 1-0 data at $\sim 3''$ resolution 
with that derived from our new $^{12}$CO J=2-1  data.  Only the fitted 
parameters are shown; the gas density and $^{12}$CO abundance are the same 
in the two analyses and do not affect the stellar mass determination because 
the $^{12}$CO J= 1-0 and J= 2-1 lines are optically thick. We scaled GD98's 
values of the parameters that depend on physical size, and hence distance, 
from 150 pc distance to the 140 pc distance used here. The
uncertainties are the formal 2$\sigma$ values from the $\chi^2$ fit. The 
position angles in Table 3 refer to the major axis of the disk; the PA 
listed in GD98 is that for the disk rotation axis so differs by 90$^\circ$ 
from the PA$_{\mathrm{CO}}$ given here. The agreement between GD98 and the 
new results is excellent for all the parameters except for the inclination 
which now has the opposite sign. It is not surprising that the higher 
resolution data better define the sense of the
inclination.  Its sign does not affect the mass determination.

Table 4 lists the parameters most relevant  for determination of the central 
mass derived from our new observations and also from our previously published
observations.  Columns 2 and 3 are described below. Columns 4 and 5 list the 
disk PA and $i$ derived from the CO line maps. The disk outer radius, 
R$_{out}$, in Col. 5 is included here because its determination  and that 
of $i$ are coupled (see, for example, GD98 Sect.4.4). Col. 6 lists the value 
of Vsin$i$ at the reference radius 100~AU and col. 7 the
exponent of the radial power law dependence of the velocity.  Within the
uncertainties, the rotation is Keplerian.  The stellar masses in col. 8 are 
derived using
\begin{equation}
M = \left\{ {{(V_{100} sin~ i)}\over{ 2.98 sin~i}} \right\}^2 ~\msun
\end{equation}
where 2.98 km/s is the circular velocity at 100 AU radius from a 1 \msun~ 
star.  The derived masses scale as $M_* \propto v^2(r) r$ and hence depend on 
distance to the star as $M_* \propto D$. The main source of uncertainty in 
the measured stellar mass is the inclination; the two objects in our sample 
with the lowest inclinations, CY Tau and BP Tau, have the highest 
uncertainties in their derived masses.  This approach ignores the mass of the 
disk because it is usually only a few per cent that of the star.  Even in
GG Tau A, which probably has the most massive disk of the stars in our
sample, the mass of the circumbinary is only $\sim 10\%$ that of the stars 
within it (Guilloteau \etal 1999).  It is interesting that UZ Tau E has a 
circumbinary ring like GG Tau A  and UY Aur (Duvert \etal 1998), although 
the central hole is as yet unresolved.

The 1.3\,mm continuum emission is detected by a separate back-end (\S 2.2)
and the detected continuum emission arises mostly from the innermost
regions of the disk (see below).  The continuum images can therefore provide 
independent measurements of the disk orientation. 
Cols.\,2-3 of Table 4 give the position angle, PA$_{cont}$ of the
major axis of the disk, and its inclination, $i_{cont}$, measured from the 
1.3 mm wavelength continuum image. To derive the inclination from the ratio 
of apparent major and minor axes, we used an effective  seeing of 0.3$''$.  
We did not include uncertainty in the seeing in the error estimate. The 
agreement between the orientations and inclinations derived from 1.3\,mm 
continuum data and $^{12}$CO J=2-1 data is excellent for most of the targets. 
The uncertainties of the values derived  from the continuum images are larger,
in most cases, than those from the CO data because the measured sizes of the 
disks in the continuum are generally smaller than in CO, specially when the 
inclinations are low.  The  sizes measured in the continuum are smaller
because the dust continuum opacity is much smaller than in the CO line.  
The detected continuum emission therefore arises 
mostly in the dense inner regions of the disks (see, for example, D96). 
The values for CY\,Tau suffer from low signal to noise. 
In the case DL\,Tau, significant confusion from the molecular cloud 
seems to have affected the determination of the inclination from the CO 
emission (\S 3). The confusion affects the channels near the systemic 
velocity, which are the most important to determine the inclination. 
Accordingly, pending better measurements (e.g. $^{13}$CO), we prefer to use 
the inclination determined from the continuum
emission, $\sim 45^{\circ}$, to derive the mass for DL\,Tau. With this 
inclination, the stellar mass is $0.72 \pm 0.11$\msun, rather than 
$1.23 \pm 0.11$\msun~ when using the CO derived inclination.

\section{Comparison of the Measured Masses and Theoretical Tracks}
\subsection{The Theoretical Tracks}
The observational measure of the stellar photospheric temperature is its 
spectral type while the parameter provided by the models is usually $T_{eff}$.
The surface gravities of the PMS stars lie  between those of the main sequence
 stars and giants so the conversion between spectral type and T$_{eff}$ 
determined for the giants and dwarfs may not apply. White \etal (1999) and 
Luhman (1999), in addition to others, have investigated this problem and 
demonstrate that the differences in  temperature scales become apparent for 
the lowest mass stars, and below the hydrogen-burning limit. Luhman (1999) 
proposed an ``intermediate'' temperature scale for the M stars. Its 
differences from  Kenyon and Hartmann's (1995) temperature scale become 
significant for stars cooler than about M6.  Since our sample does not 
include stars as cool as this, we use the Kenyon and
Hartmann scale in the present work without modification.

The HR diagram in Fig. 2a shows excerpts from  PMS tracks  calculated by 
DM97, BCAH, PS99 and SDF.  In this and subsequent figures, the BCAH tracks 
for $M/\msun < 0.7$ are for mixing length parameter (mixing length/pressure 
scale height)= 1.0 and 1.9 for $M/\msun \ge 0.7$ (BCAH98 and Baraffe, priv. 
comm.). Differences among the theoretical calculations for a star of given 
age and mass are obvious.  For example, the DM97 tracks are hotter than those 
of BCAH for $M_* \sim 0.2 ~\msun$.  SDF's models at 0.2 and 0.1 \msun
~appear to contract more slowly than those of DM97, BCAH, and PS99. Since the 
different treatments of stellar convection, the equation of state, opacities, 
and stellar atmospheres become important in specific regions of the HR 
diagram, tests of the theoretical tracks require accurate measurements of 
mass over as wide a range of mass and age as possible.

The dependence of $M_*$ on distance (\S 3) is important because while the 
average distance to the Taurus star forming region (SFR) is reasonably well 
known, the actual distance to a given star in it is not.  The Taurus SFR 
extends for at least $15^{\circ}$ on the sky.  If its depth is comparable to 
its width, the distance to a specific member may scatter by $\pm 20$ pc 
around the 140 pc mean, a relative error of $\pm14\%$.  Only CY Tau and 
BP Tau (Table 4) have a larger uncertainty of their mass measurement. For 
most of the stars in Table 4 the uncertainty in their actual mass is 
therefore dominated by the uncertainty in their distance.  The distance of
course also determines the star's location in the HR diagram because 
$L_* \propto D^2$. We will therefore present most of our results in \S4 
on modified HR diagrams in which the distance-independent parameter $L/M^2$ 
is plotted versus $T_{eff}$. For reference, Fig. 2b compares the tracks 
on this basis.

\subsection{Measured Masses on Theoretical HR Diagrams: The Single Stars}

Fig. 3 plots the single T Tauris in our sample on modified HR diagrams
using PMS tracks calculated by DM97, BCAH, PS99, and SDF. The uncertainties 
displayed along the horizontal axis are $\pm 1$ spectral type sub-class.  
Along the vertical axis, the uncertainties are the propagated internal 
uncertainties in stellar mass (Table 4) and an assumed $\pm 10\%$ uncertainty 
in the luminosity given in Table 1. We discuss the fit of each star to the 
theoretical tracks separately.

\paragraph{Lk\,Ca\,15:}  The $L/M^2$ value for Lk\,Ca\,15 lies near the 
1.0 \msun ~tracks calculated by BCAH, PS99, and SDF (Fig. 3) consistent 
with its mass $0.97\pm0.03$ \msun~ at the nominal 140 pc distance (Table 4).  
These tracks yield a consistent age estimate of 3-5 MY.  
Lk\,Ca\,15's $L/M^2$ values  lies on the 0.8 \msun ~track calculated by DM97.
Agreement with the DM97 track would require a distance of $\sim 115$ 
pc but would not affect the age estimate.

\paragraph{DL Tau:} The position of DL Tau's $L/M^2$ in the diagrams is 
within one spectral type uncertainty of the 0.8 \msun ~track for each of 
the calculations. This is in good agreement with the mass derived using 
the continuum inclination, $0.72 \pm 0.11$ \msun.  The theoretical tracks 
yield a consistent age estimate of $\sim 2$ MY.

\paragraph{GM Aur:}  The position of the $L/M^2$ value relative to the BCAH,
PS99, and SDF tracks is consistent with its mass $0.84\pm0.05$ \msun~at 
140 pc. The tracks provide an age estimate of $\sim 3$ MY.  The $L/M^2$ 
value lies close to DM97's  0.6 \msun ~track which would require a distance 
of 103 pc to force agreement with the dynamical mass.

\paragraph{BP Tau:}  BP Tau's $L/M^2$ values lies between DM97's 0.6 and 
0.8 \msun ~tracks, and close to BCAH, PS99, and SDF's 0.8 \msun ~tracks, 
while our measured mass at 140 pc is in the range 0.92 to 1.49 \msun.  This 
suggests that the distance to BP Tau may be closer than 140 pc but not as 
extreme as the \textit{HIPPARCOS} value, 53$^{+17}_{-11}$ pc (Favata \etal 
1998).  The large uncertainty in the mass produces a large spread in the 
age estimate, 2-10 MY.

\paragraph{DM Tau:}  Its dynamical mass at 140 pc, $0.55\pm0.03$ \msun, is 
consistent with its $L/M^2$ value relative to four sets of tracks.  The 
tracks yield a consistent age estimate $\sim 5$ MY.

\paragraph{CY Tau:}  The nominal value 0.55 \msun ~is
consistent with its $L/M^2$ relative to the four sets of tracks with an 
age in the range 2-5 MY.  However, the uncertainty in its mass is so large 
that this agreement is not significant, and we cannot plot it meaningfully 
in Fig. 3.

\paragraph{MWC480 (=HD 31648):} Fig. 4 provides a similar comparisons of 
the dynamical mass of the HAeBe star MWC480.  We calculated its stellar 
luminosity (Table 1) for a distance of 140 pc using the spectral energy 
distribution, corrected for extinction and circumstellar emission, derived 
by Malfait \etal (1998). This value, 11.5 \lsun, is about half that 
calculated by Mannings and Sargent (1997) from photometry available at 
the time.  Table 1 also lists Grady's (1999) spectral type estimate, A4,
from HST spectra, which is cooler than Th\'e et al's (1994) earlier 
value, A2-A3.  BCAH PMS models do not extend to $M > 1.2 ~\msun$, so we 
make the comparisons only with DM97, PS99, and SDF's tracks.  MWC480's  
$L/M^2$ value lies close to the 2.0 \msun ~tracks for the three calculations
at an age $\sim 7$ MY.  The mass at 140 pc distance $1.65\pm0.07 \msun$ 
suggests that MWC\,480 lies at a somewhat greater distance.  A
distance of 170 pc would yield a dynamical mass of 2.0 \msun and  
would be within $2 \sigma$ of the \textit{HIPPARCOS} measurement of 
$131^{+24}_{-18}$ pc (van den Ancker et al 1998).

\subsection{The Binaries}
The results for the binaries cannot be plotted on $L/M^2 ~vs ~T_{eff}$ diagrams
because the component luminosities of UZ Tau E or component masses of 
GG Tau are not known.   We discuss the binaries using conventional HR diagrams.

\paragraph{UZ Tau E:}  Mathieu et al (1996) discovered that UZ Tau E is a 
single-lined spectroscopic binary with period 19.1 days and projected 
semi-major axis of the primary $a_1~sin~i_* = 0.03$ AU in which $i_*$ is 
the inclination of the orbit.

Our measurement of the total binary mass $(M_1 + M_2)$ (Table 4) and the period
determine $(a_1 + a_2) = (0.153\pm0.003) (D(pc)/140)^{1/3}$ AU.  We can 
solve for $a_1$ and $a_2$, and hence $M_1$ and $M_2$ if we assume that the 
circumbinary disk and binary orbit are coplanar, $i_{CO} = i_*$.  This 
assumption yields $M_1 = (1.00\pm0.08)(D(pc)/140)$\msun~ and $M_2 = 
(0.31\pm0.02)(D(pc)/140)$ \msun. We plot
the primary in the HR diagram (Fig. 5) assuming that the system spectral 
type M1 (Table 1) applies to the primary, assigning, as before, an 
uncertainty of $\pm 1$ one sub-class, and luminosity in the range 
0.5 to 1 times the total system luminosity, 1.6 \lsun.  Fig. 5 shows that, 
with these assumptions, the primary lies in the $\sim 0.6$ \msun~region 
of the HR diagram for all the tracks, clearly discrepant with the derived 
mass, $\sim 1$ \msun. It seems very unlikely that the distance to UZ Tau E 
could be  $\sim 60\%$ of the nominal 140 pc.  Rather, the assumption that 
$i_* = i_{CO}$ is probably wrong; the circumbinary disk and binary star 
orbits appear not to be coplanar.  The spectroscopic binary orbit will be 
resolvable by the next generation space-borne astrometric missions so a 
precision measurement  of the orbital parameters should be possible within a 
decade.

\paragraph{GG Tau:}  White \etal (1999) compared the (L,T$_{eff}$) 
positions of GG Tau Aa and Ab, the components of the $\sim 0.25''$ separation 
binary at the center of the circumbinary ring we observed, with PMS 
evolutionary tracks calculated by Swenson \etal (1994), D'Antona and 
Mazzitelli (1994), DM 97, and BCAH.  They used the effective temperature 
scales for main sequence dwarfs drawn from the works
of Bessel (1991), Leggett (1996), and Luhman and Rieke (1998).  They 
found that the tracks calculated by BCAH were the most consistent with 
the total dynamical system we have measured and the expected coevality of 
the components. The BCAH indicated an age between 1 and 2 MY.

Fig. 6 shows a similar comparison with the tracks of DM97, BCAH, PS99, and 
SDF. The difference between the temperature scale used by White 
et al and Kenyon and Hartmann's (1995) temperature scale is unimportant at the 
effective temperatures of GG Tau Aa and Ab. The BCAH tracks  suggest primary 
and secondary masses of $\sim 0.8$ and $\sim 0.7$ \msun, and age 1-2 MY, in 
agreement with White et al's finding. The PS99 and SDF tracks suggest similar 
masses and age. The DM97 tracks also indicate a young age for the
components, in this case less than 1 MY.  The total mass indicated by DM97's 
tracks, $\sim 0.95$ \msun, is significantly lower than our measured 
value. To reconcile it with the measured mass would require that the GG Tau 
system lie at distance $\sim 103$ pc which seems unlikely.

\section{Summary and Suggestions for Future Work}

Our results indicate that:

1) Stellar masses  can be measured dynamically by the
rotation of their circumstellar disks with high enough precision, $<5\%$,
that meaningful tests of calculations of pre-main sequence  evolution are 
possible.

2) The BCAH, PS99, and SDF models in general are in reasonable agreement with 
the measured dynamical masses at the average distance to the Taurus SFR. To 
force agreement between the DM97 models and mass measurements for stars in the 
$\sim 0.7$ to 1  \msun ~range (Lk\,Ca\,15, GM\,Aur, and the components of 
GG\,Tau) would require that these stars lie at distances 100-115 pc which 
seem unacceptably near. This is a consequence of the apparently warmer 
photospheres of DM97's models for stars close to 1 \msun.

3)  Consistency of the position of the HAeBe star MWC\,480 with respect to 
DM97, PS99, and SDF's tracks suggest it lies on the far side of the Taurus 
SFR at $\sim 170$pc.

Our tests of the pre-ms tracks are limited by our present ignorance of actual
distances to individual stars.  We anticipate that this limitation will be 
overcome
by precision distances that will be measured by the astrometric observatories
currently under construction (e.g. FAME, GAIA, and SIM).  All our measurements 
at present pertain to stars with $M>0.5 \msun$ so we have been  unable to test 
the models at the lowest end of the
stellar mass spectrum.  This is a serious limitation and results, at least in 
part, from the range of masses of stars formed in the Taurus SFR.  
We look forward to the capabilities  of the ALMA mm-wave interferometric 
array which will open the rich Orion SFR to investigation. Conversion of the 
observationally derived parameter,
spectral type, to that provided by the models, $T_{eff}$, is a
lingering source of uncertainty, particularly at the lowest masses. Since stars
contracting to the main sequence represent a range of surface gravities, the
conversion may depend on the stars age. Relief from this problem should become
available soon as the model calculations which include stellar atmospheres 
provide diagnostics such as color indices and model spectra, for closer 
comparison with the observations.  We look forward to a fruitful decade ahead.

\acknowledgements We thank I. Baraffe, J. Bouvier, G. Chabrier, C. Dougados, M.
Forestini, F. Palla, S. Stahler, and L. Siess for helpful conversations and
communicating results before publication. This work began as a collaboration 
with G. Duvert and F. M\'enard; we thank them for their advice and help in 
its early stages.  We thank the referee for a prompt and helpful report.
This paper was begun during M.S.'s sabbatical 
visit at the Observatoire de Grenoble; M.S. thanks G. Duvert and C. Perrier 
for making it possible and for their hospitality.  The work of M.S. was 
supported in part by NSF Grants 94-17191 and 98-19694.

{\it In memoriam:}  The Plateau de Bure interferometer was affected 
by two transportation tragedies in 1999.  These caused the death of 25 
individuals including six members of the IRAM staff and two from INSU-CNRS. 
We dedicate this work to the memory of the victims, especially our 
colleagues B.Aubeuf, F.Gillet, H.Gontard, D.Lazaro, R.Prayer and P.Vibert.

\begin{deluxetable}{llllll}
 \tablecolumns{6}
 \tablewidth{0pt}
 \tablecaption{Parameters of Program Stars \label{table1}}
 \tablehead{
 \colhead{Name}  & \colhead{SpT}  &
 \colhead{L$_*/\lsun$} & \colhead{T$_{eff}$(K)} & \colhead{Refs} &
 \colhead{Refs}\\
 \colhead{} & \colhead{} & \colhead{} & \colhead{} & \colhead{(CO)} 
& \colhead{(L$_*$, T$_{eff}$)}} 
 \startdata
 Singles:  &        &      &             &   &      \\
 MWC\,480   &  A4    & 11.5~&      8460   & 1 & 1  \\
 Lk\,Ca\,15    &  K5    & 0.74 &      4350   & 2 &   \\
 DL Tau    &  K7    & 0.68 &      4060   & 3 & 2  \\
 GM Aur    &  K7    & 0.74 &      4060   & 4 & 3  \\
 BP Tau    &  K7    & 0.93 &      4060   &   & 4 \\
 DM Tau    &  M1    & 0.25 &      3720   & 5 &    \\
 CY Tau    &  M1    & 0.47 &      3720   & 3 &  \\
 Binaries: &        &      &             &   &  \\
 GG Tau Aa &  K7    & 0.84 &     4060    & 6 & 5   \\
 GG Tau Ab &  M0.5  & 0.71 &     3800    &   &   \\
 UZ Tau E  &  M1    & 1.60 &     3720    & 3,7&  \\
 \enddata
\tablecomments{References for Col. 5:  1- Mannings and Sargent, 1997, 2- Duvert
\etal 2000, 3- Dutrey \etal 1996 (D96), 4-Dutrey \etal 1998, 5-GD98, 
6-Guilloteau, Dutrey and Simon 1999, 7-Jensen, Koerner, and Mathieu 1996 
References for Col. 6: 1) L$_*$ from Malfait \etal (1998) and this work, 
spectral type from Grady (1999), 2) L$_*$ from Hartigan \etal (1995),
 3) Spectral type and L$_*$ from Gullbring \etal (1998) 4) See Favata 
\etal, 1998 and Sect. 4.1, 5) L$_*$ and spectral types
from White \etal (1999).}
\end{deluxetable}

\begin{deluxetable}{lllllll}
 \tablecolumns{7}
 \tablewidth{0pt}
 \tablecaption{Measured mm Continuum Source Coordinates (J2000.0)\label{table2}}
 \tablehead{\colhead{Name} & \multicolumn{3}{c}{RA}& \multicolumn{3}{c}{Dec}}
 \startdata
 MWC\,480   &04$^h$ &58$^m$ &46.264$^s$ &29$^\circ$ &50$'$ &37.03$''$ \\
 Lk\,Ca\,15  &04 &39 &17.784 &22 &21 &03.46   \\
 DL Tau    &04 &33 &39.074 &25 &20 &38.14   \\
 DM Tau    &04 &33 &48.731 &18 &10 &10.00   \\
 CY Tau    &04 &17 &33.727 &28 &20 &46.95   \\
 UZ Tau E  &04 &32 &43.070 &25 &52 &31.14   \\
 BP Tau    &04 &19 &15.835 &29 &06 &26.90 \\
 \enddata
\tablecomments{The astrometric accuracy is better than $0.05''$.}
\end{deluxetable}

\begin{deluxetable}{llcc}
 \tablecolumns{4}
 \tablewidth{0pt}
 \tablecaption{Comparison of Old and New Results for DM Tau \label{table3}}
 \tablehead{ \multicolumn{2}{c}{ Parameter} & GD98 & This Work}
 \startdata
 Systemic velocity & $V_{LSR}$ (km/s      & $6.05\pm 0.02$ &$6.01\pm0.01$\\
 Orientation       & PA$_{\mathrm{CO}}$ (deg)      & $-25\pm2    $ &$-23\pm1$\\
 Inclination       & $i$ (deg)            & $33\pm2 $      &$-32\pm2$\\
 Outer radius        & $R_{out}$ (AU)     & $793\pm20   $  &$800\pm5$\\
 Turbulent linewidth  & $\Delta v$ (km/s) & $0.08\pm0.03$  &$0.07\pm0.02$\\
 \tableline
 \multicolumn{4}{c}{ } \\
 \multicolumn{4}{c}{ Temperature law:
 ~~~ $T(r) = T_{100} (\frac{r}{100 \rm{AU}})^{-q}$  }              \\
 \multicolumn{4}{c}{ } \\
      temperature at 100 AU & $T_{100}$ (K) & $31\pm2 $ &$32.5\pm0.5$\\
      temperature exponent  & $q$       & $0.63\pm0.05$ &$0.63\pm0.01$\\
 \tableline
 \multicolumn{4}{c}{ } \\
 \multicolumn{4}{c}{Velocity law: $V(r) = V_{100} (\frac{r}{100 {\rm AU}})^{-v}$}   \\
 \multicolumn{4}{c}{ } \\
 V sin$i$ at 100 AU& $V_{100}$sin$i$ (km/s) & $1.11\pm0.03$ & $-1.17\pm0.02$ \\
 velocity exponent& $v$                     & $0.57\pm0.04$ & $0.53\pm0.01$ \\
 stellar mass              & M$_*$ ($\msun$)& $0.47\pm0.06$ & $0.55\pm0.07$
 \\
\enddata
\tablecomments{Comparison of parameters derived from CO(1-0) (GD98) and CO(2-1)
(this work) in DM Tau.}
\end{deluxetable}

\begin{deluxetable}{lrlrllrcc}
\tabletypesize{\small}
 \tablecolumns{9}
 \tablewidth{0pt}
 \tablecaption{Derived Parameters (D=140 pc) \label{table4}}
 \tablehead{ & \colhead{PA$_{cont}$} & \colhead{$i_{cont}$} & \colhead{PA$_{CO}$} &
 \colhead{$i_{CO}$} & \colhead{$R_{out}$} & \colhead{$V_{100} sin(i)$} & \colhead{$-v$}
 & \colhead{M$_*$} \\
\colhead{}& \colhead{deg} &\colhead{deg} & \colhead{deg} & \colhead{deg} & 
\colhead{AU} & \colhead{km/s} &\colhead{ } &  \colhead{\msun} \\
  }
\startdata
\multicolumn{9}{c}{Singles}\\
 MWC\,480   &$ -10 \pm11$ & $26 \pm7$&$-32\pm1$&$+38\pm1$&$545\pm5$
  &$-2.38\pm0.02$& $0.50 \pm 0.02$            &$1.65\pm0.07$\\
 Lk\,Ca\,15    &$62 \pm5$&$42 \pm 5$    & $+61\pm1$&$+52\pm 1$&$ 650\pm15$
  &$-2.30\pm0.02$&$0.56\pm0.03$&$0.97\pm0.03$\\
 DL Tau$^1$      &$44 \pm3$& $49 \pm3$& $+50\pm3$&$+35\pm 2$&$ 520\pm50$
  &$ 1.90\pm0.06$&$0.55\pm0.03$& $0.72\pm0.11$ \\ 
 GM Aur      &$57 \pm 5$& $54 \pm5$&  $+51\pm2$&$+56\pm 2$&$ 525\pm20$
  &$ 2.30\pm0.08$&$0.5\pm0.1  $&$0.84\pm0.05$\\
 DM Tau    & $-1 \pm5$ & $45 \pm5$& $-27 \pm 1$ &$-32\pm1 $&$ 800\pm5 $
  &$-1.17\pm0.02$&$0.53\pm0.01$&$0.55\pm0.03$\\
 CY Tau    &$-56 \pm7$&$47 \pm 8$ &     $-30 \pm7$     &$+30\pm10$&$270
\pm10$
  &$-1.10\pm 0.10  $&  $0.50 \pm 0.08$      &$0.55\pm0.33$\\
 BP Tau    & - & $20 \pm 20$ & $-28 \pm 3$ & $+30~^{+4}_{-2}$  &
$108\pm4$
  &$1.67\pm0.06$ & $0.54\pm0.07$ &$ 1.24~^{+0.25}_{-0.32}$\\
\multicolumn{9}{c}{Binaries}\\
 GG Tau A  & $7\pm2$&$37\pm1$ & $~+7\pm2$  &$+37\pm1 $&$\sim 800 $
   &$2.05\pm0.06$ &$0.5\pm0.1$  &$1.28\pm0.07$\\
 UZ Tau E     &$87 \pm3$& $54 \pm 3$ &$+86\pm2$ &$-56\pm2 $&$ 300\pm20$
   &$-2.83\pm0.05$&$0.53\pm0.03$&$1.31\pm0.08$\\
\enddata
\tablecomments{No sign for $i_{cont}$, 1- Using the inclination
derived from the continuum (see text).}
\end{deluxetable}

\newpage

\figcaption{{\it a)} Results of our interferometric  observations of   
MWC 480, DM Tau,  LkCa15, and GM Aur. The coordinates are arc-sec in RA 
and Dec centered on the peaks of the continuum emission (Table 2).  The 
long and short arms of the crosses, centered on these positions, 
indicate the apparent major and minor axes of the disks.  The long
arms are scaled to a full length of 4 arc-sec and the short arms are
foreshortened according to the inclinations.  The long arms point in the
direction of PA$_{CO}$ (Table 4). The top 7 panels for each object are velocity
channel maps with the velocity (km/sec) with respect to the local standard 
of rest indicated in the upper left corner. The contour spacing, different
for each object, is indicated in upper right corner of the top panel.  The
contour spacing is 2.7 $\sigma$ on average but varies from source to
source; the smallest value is $2.2 \sigma$ for DL Tau in Fig. 2. 
The bottom panel shows the velocity
gradient map for each object, color-coded so that red indicates red-shift
with respect to the systemic velocity.  The velocity contour spacing 
(km/sec) in these panels is: MWC 480, 0.3; DM Tau, 0.2; LkCa 15, 0.3;
GM Aur, 0.3. {\it b)}  Same as Fig. 1a but for UZ Tau E, CY Tau, DL Tau, 
and BP Tau.  The nearly emission-free channels at V$_{lsr}$ = 6.6 km/sec 
(UZ Tau) and V$_{lsr}$ = 5-6.6 km/sec (DL Tau) are the result of 
confusion with the molecular cloud.  The velocity contour spacing for the 
velocity gradient maps in the bottom panels are, in km/sec: UZ Tau, 0.6; 
CY Tau, 0.2; DL Tau, 0.4; BP Tau, 0.3.\label{fig1}}

\figcaption{{\it Top:} An HR diagram showing theoretical evolutionary 
tracks for stars of mass 0.2, 0.6, and 1.0 \msun~ calculated by DM97, 
BCAH, PS99, and SDF between the ages 2 MY  (filled circles) and 10 MY 
(open circles). {\it Bottom:} The same tracks as in upper panel plotted 
as $L/M^2$ vs $T_{eff}$ (see text).  The spectral type-effective 
temperature conversion is that of Kenyon and Hartmann (1995), see text.
\label{fig2}}

\figcaption{{\it Upper Left:} The results for the single T Tauri stars plotted 
as $L/M^2$ vs $T_{eff}$ for theoretical evolutionary tracks for 
stars of mass 0.1, 0.2, 0.3, 0.4, 0.5, 0.6, 0.7, 0.8, 0.9, 1.0 and
1.2 \msun~ calculated by DM97. The filled dots are at ages 1, 2, 3, 5, 7, 
and 10 MY. {\it Upper Right:} Same, as but for tracks calculated by BCAH. 
Here the indicated ages are 2, 3, 5, 7, and 10 MY. {\it Lower Left:}
Same for tracks calculated by PS99 at masses 0.1, 0.2, 0.4, 0.6, 0.8, 1.0, 
and 1.2 \msun. The indicated ages 1, 2, 3, 5, and 10 MY.{\it Lower Right:}
Same for tracks calculated by SDF.  The range of masses and indicated
ages are same as for DM97.\label{fig3}} 

\figcaption{{\it Top:} $L/M^2$ vs $T_{eff}$ for the HAeBe star MWC480 plotted 
with DM97 tracks for stars of mass 1.0, 1.2, 1.5, 2.0 and 2.5 \msun. The dots 
indicate ages 1, 2, 3, 5, 7, and 10 MY. For the 2.5 \msun~track, the isochrone 
dots stop at 7 MY at this age the star is on the main sequence.{\it 
Middle:}  Same but for tracks calculated by PS99.  The indicated ages are 
1, 2, 3, 5, and 10 MY.   The 2.5 \msun~star is on the main sequence at 5 MY.
{\it Bottom:}  Same but for SDF tracks. The range of masses and indicated
ages are same as for DM97.\label{fig4}}

\figcaption{The primary of the spectroscopic binary UZ Tau E on an HR diagram 
with evolutionary tracks plotted in the same format as in Fig. 3.  The 
system spectral type M1 is assumed to apply to the primary with an uncertainty 
of $\pm1$ sub-class.  Its luminosity is plotted in the range from 
total system luminosity to half this value.  If the inclinations of the 
circumbinary CO disk and the stellar binary are assumed the same, the mass 
of the primary is $1.0\pm0.08~$\msun~ (see text).\label{fig5}}

\figcaption{{\it Upper Left:} The components of GG Tau, Aa and Ab, plotted on 
HR diagrams showing the DM97 tracks.  The masses and indicated ages are the 
same as in Fig. 3. {\it Upper Right:} Same but using BCAH tracks, presented in 
same format as corresponding panel in Fig. 3. {\it Lower Left:}  Same 
but for PS99 tracks, in same format as corresponding panel Fig. 3.{\it 
Lower Right:} Same, but for SDF tracks, following format of Fig. 3. 
\label{fig6}}

\vfill\eject
\epsfbox{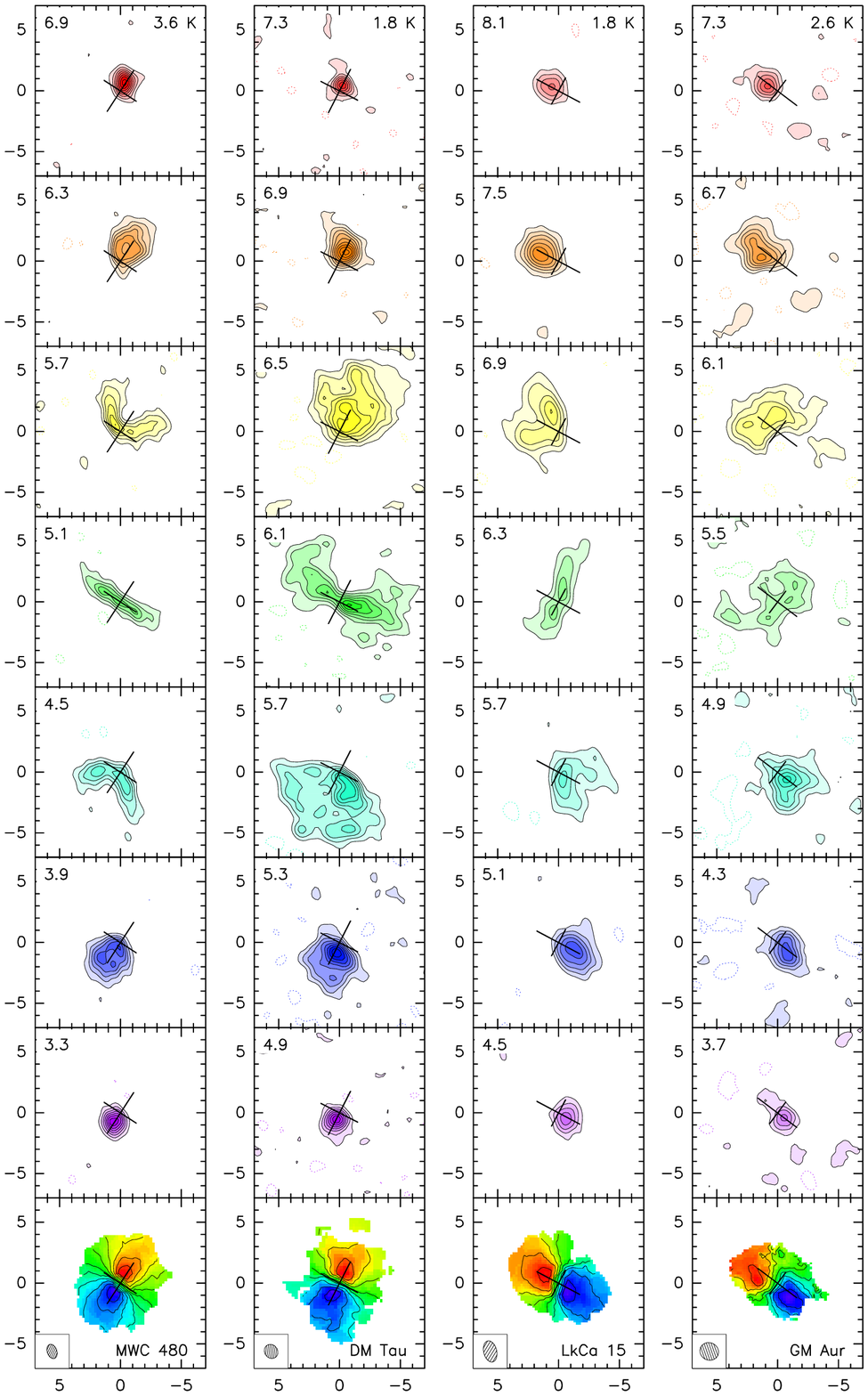}
\vfill\eject
\epsfbox{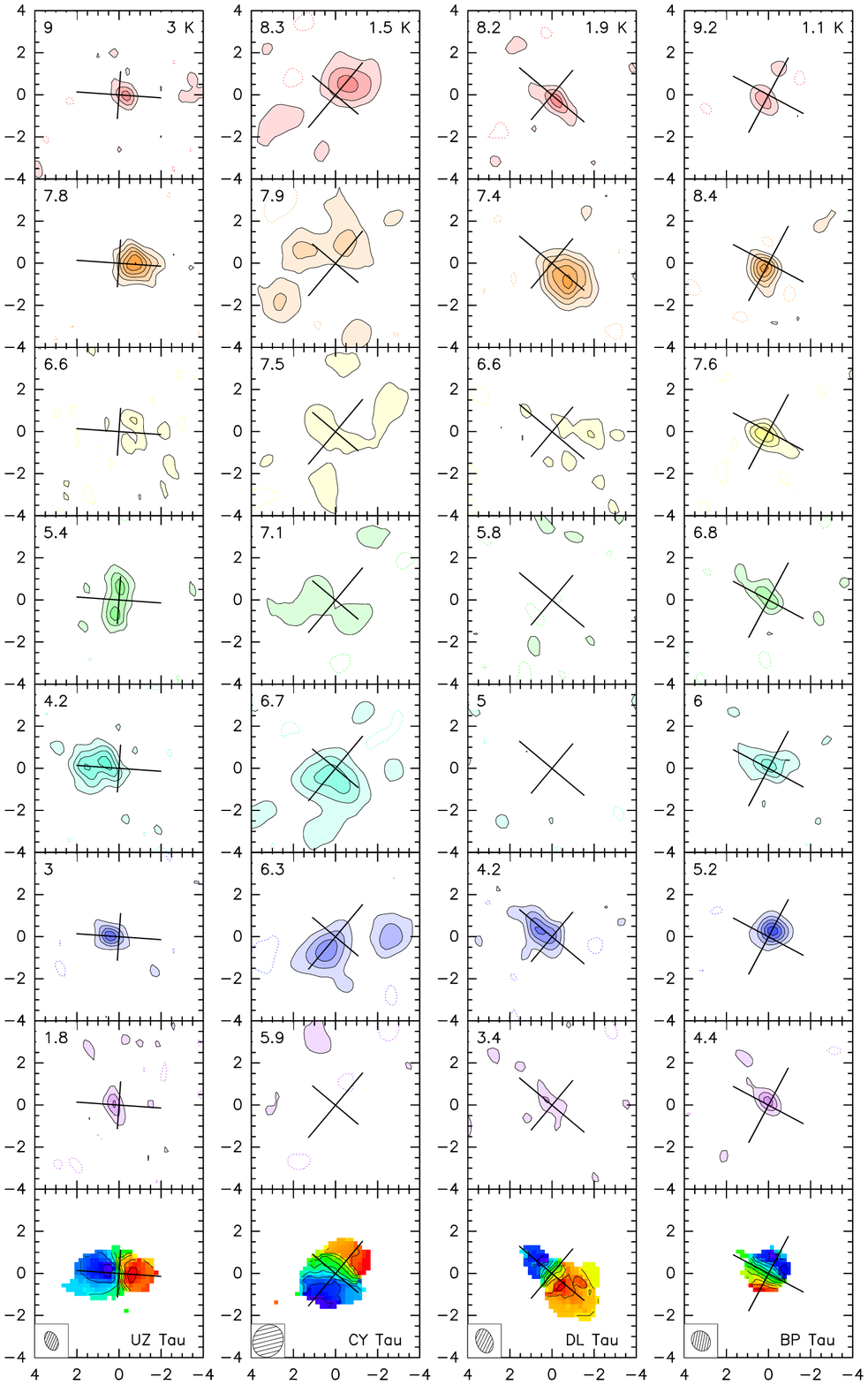}
\vfill\eject
\epsfbox{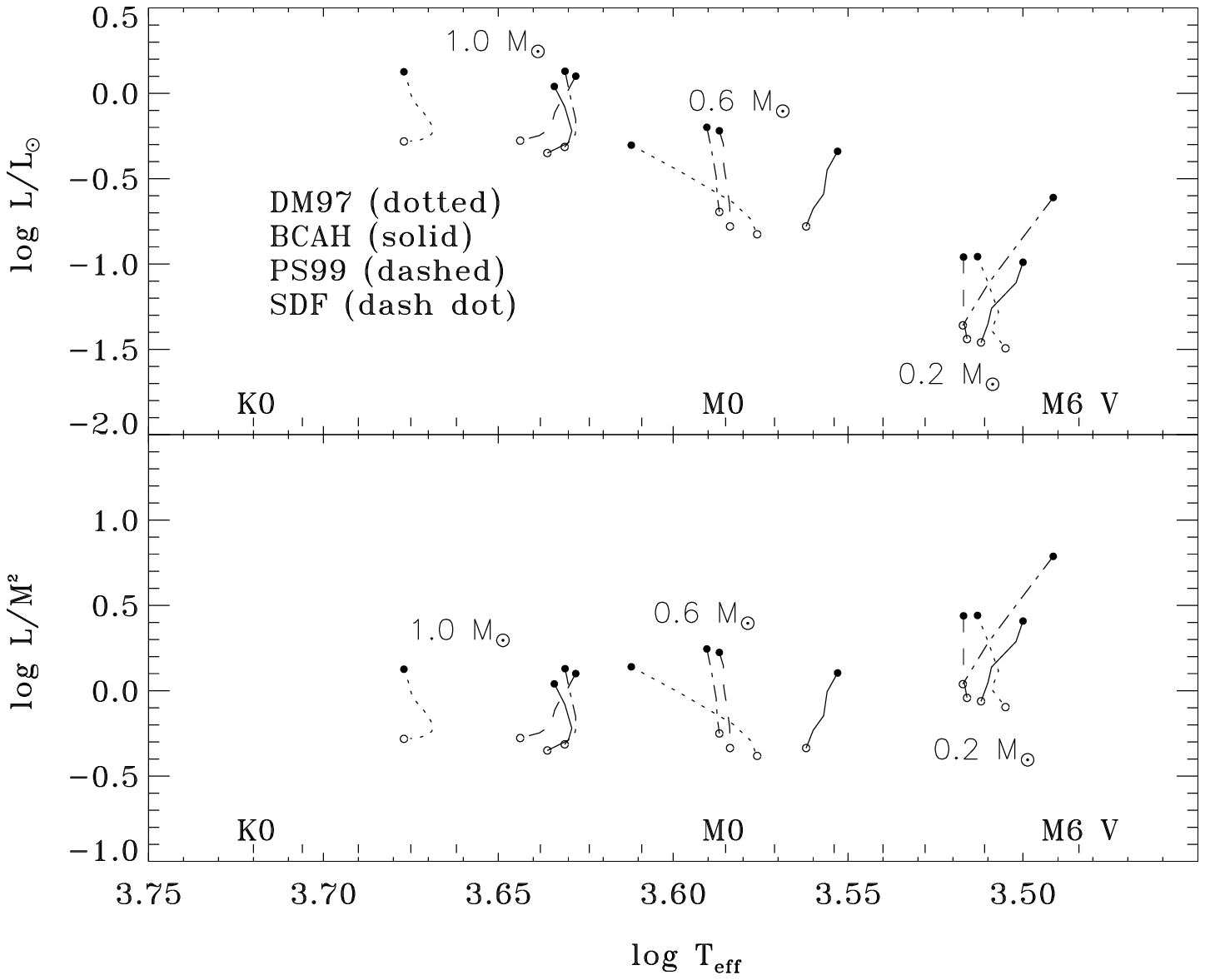}
\epsfbox{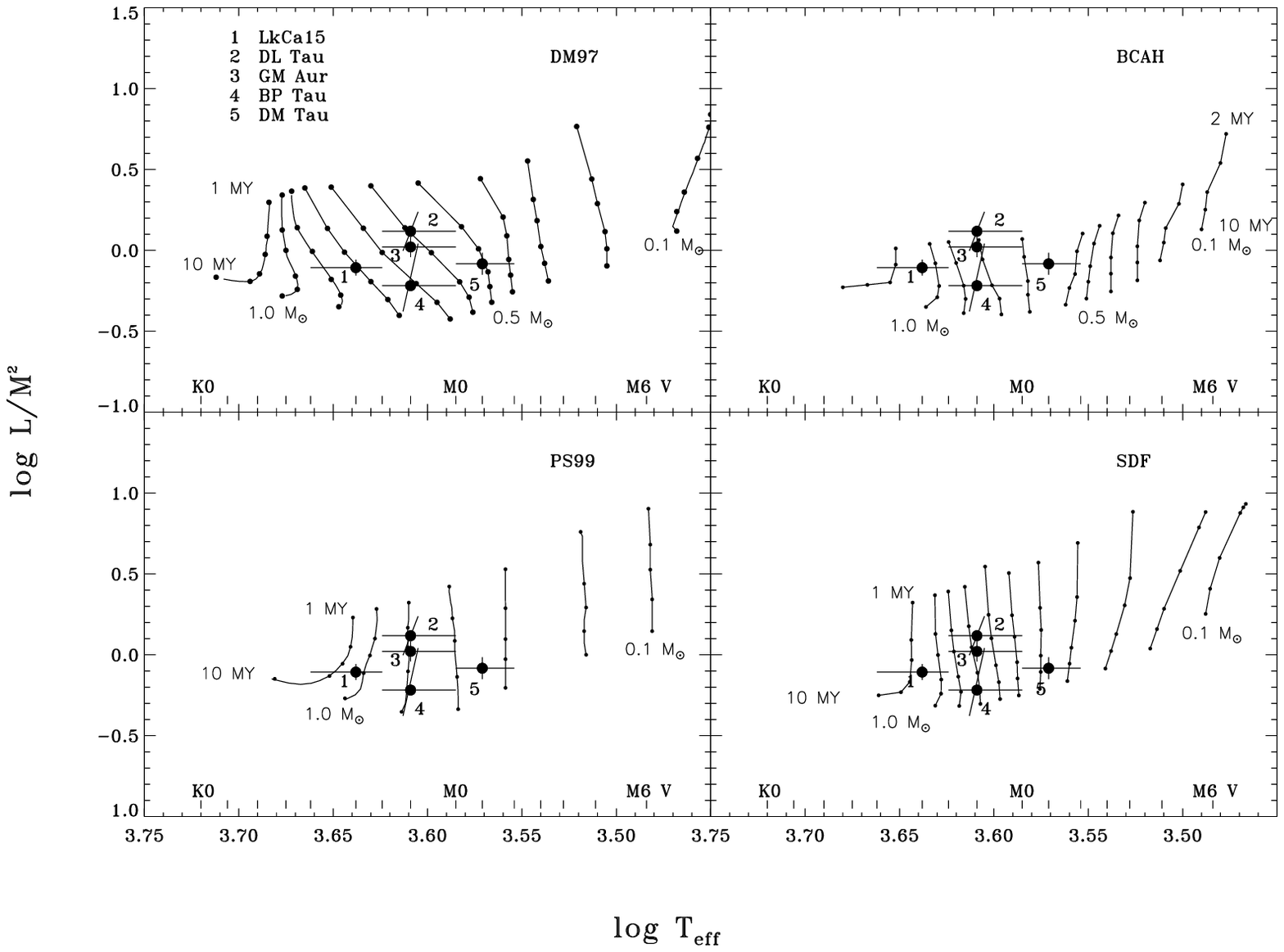}
\epsfbox{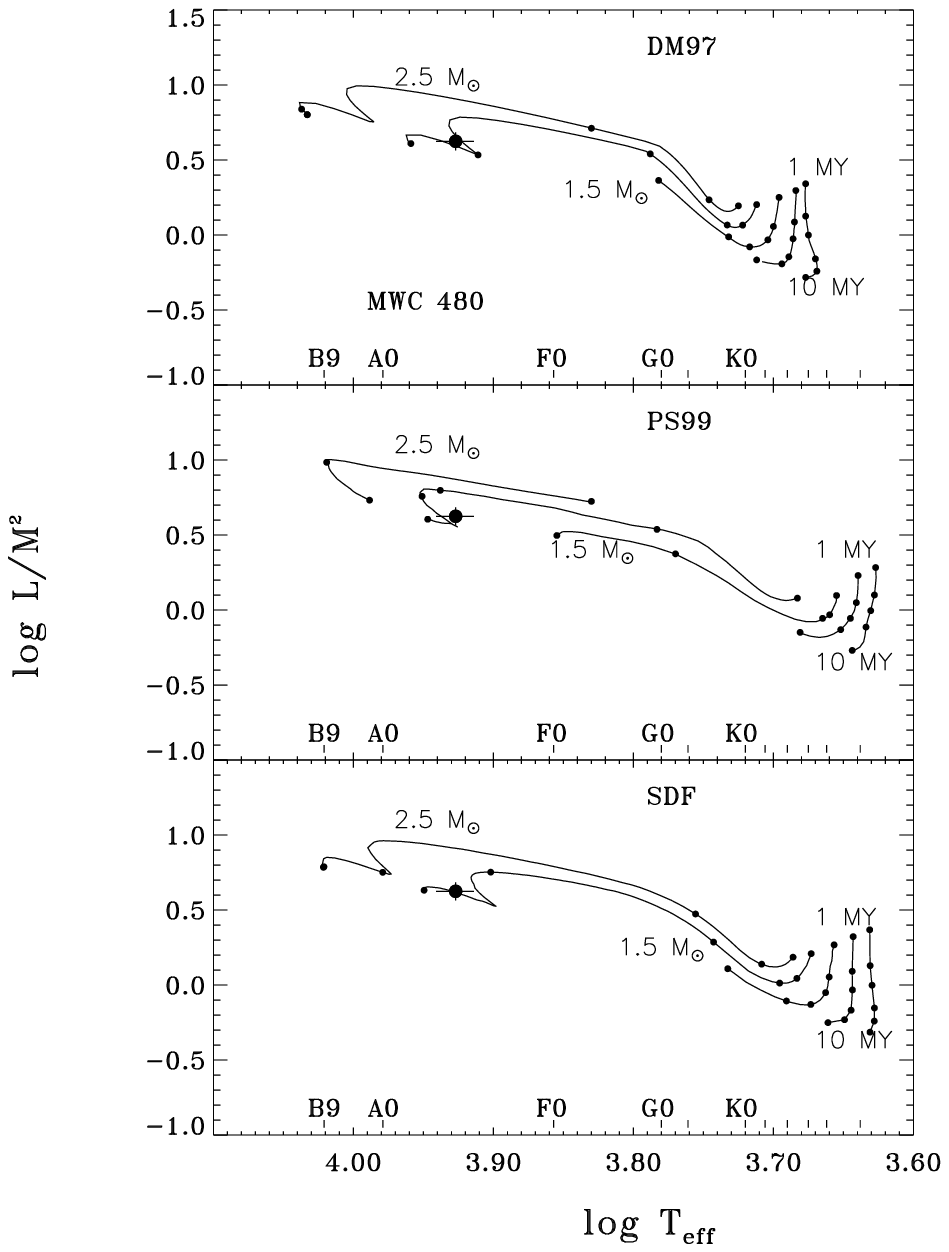}
\epsfbox{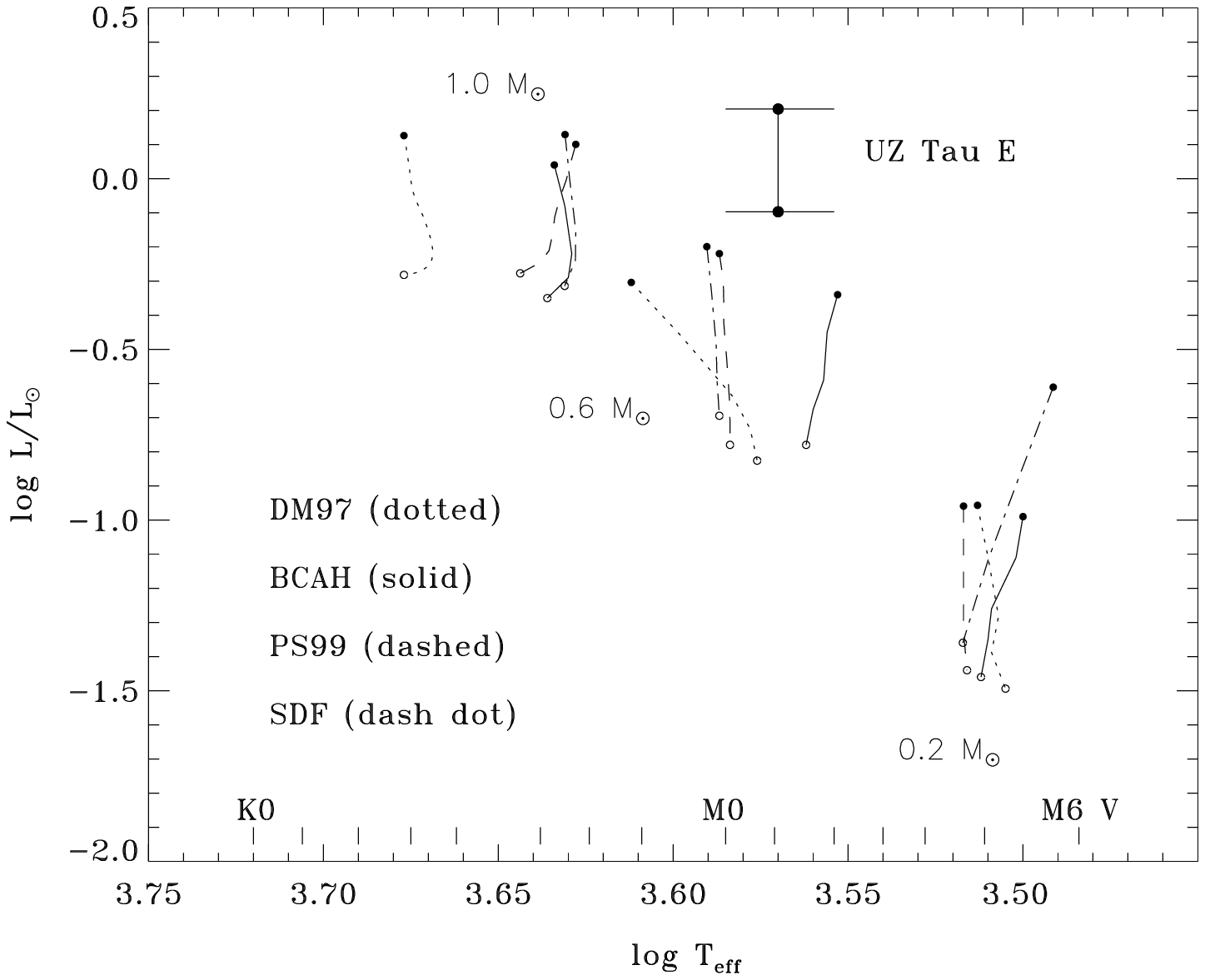}
\epsfbox{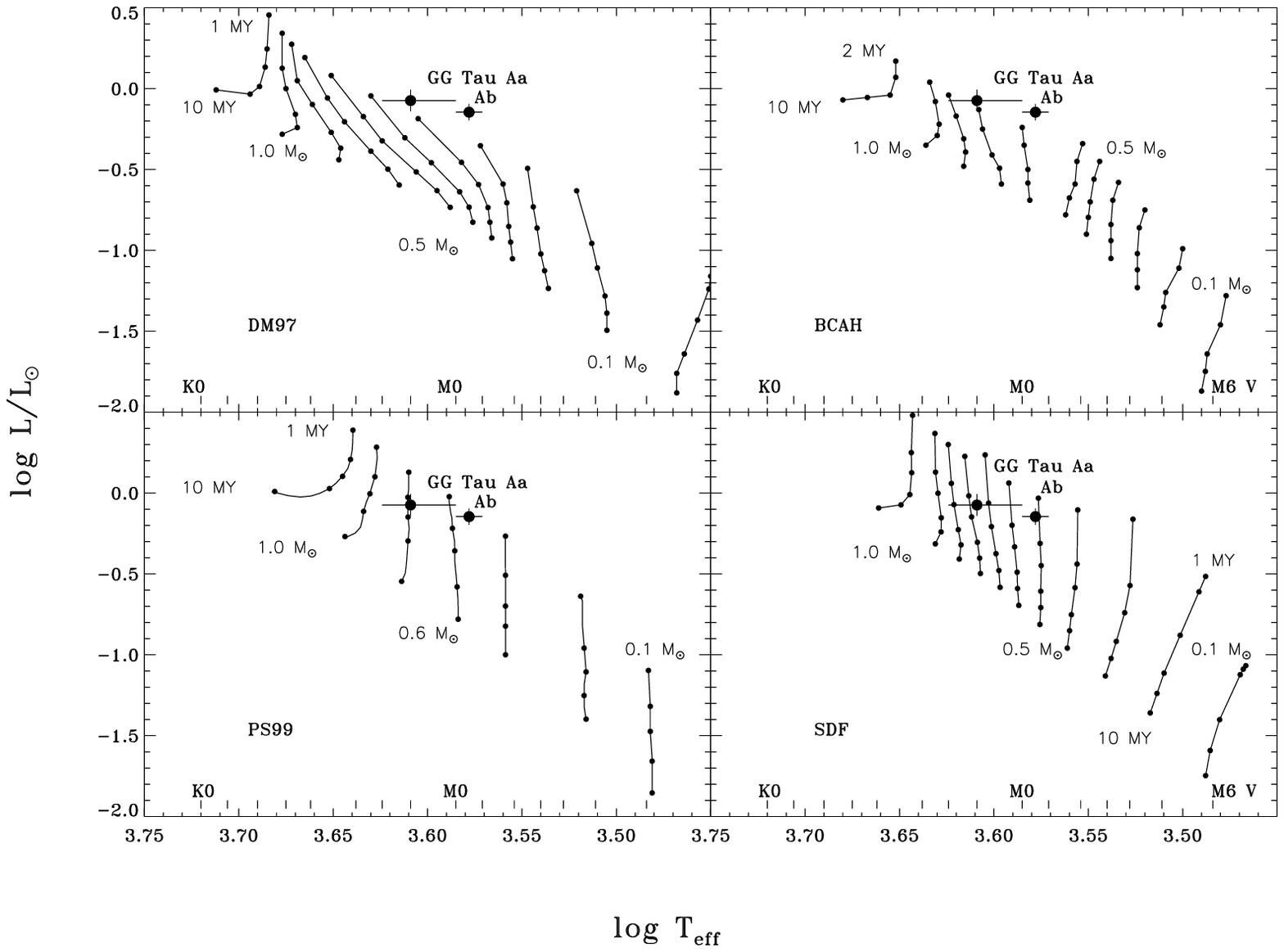}

\end{document}